\documentclass[twocolumn,showpacs]{revtex4}

\usepackage[dvips]{graphicx}%
\usepackage{dcolumn}
\usepackage{amsmath}
\usepackage{epsfig}

\begin{document}

\preprint{gr-qc/0108054}

\title{A Conventional Physics Explanation for the Anomalous
Acceleration of Pioneer 10/11}% Force line breaks with \\

\author{Louis K. Scheffer}
 \email{lou@cadence.com}
\affiliation{%
Cadence Design Systems \\
555 River Oaks Parkway \\
San Jose, CA 95134
}%

\date{\today}% It is always \today, today, but you may specify any date with \date.

\begin{abstract}
Anderson, {\it et al.}\ find the measured trajectories of Pioneer 10 and 11 
spacecraft deviate from the trajectories computed from known forces acting
on them.  This unmodelled acceleration can be accounted for by
non-isotropic radiation of spacecraft heat.  Various forms of non-isotropic
radiation were proposed by Katz, Murphy, and Scheffer, but Anderson, 
{\it et al.}\ felt that none of these could explain
the observed effect.  This paper calculates the known effects in more detail and
considers new sources of radiation, all based on spacecraft construction.
These effects are then modelled over the duration of the experiment.
The model provides a reasonable fit to the
acceleration from its appearance at a heliocentric distance of 5 AU to the
last measurement at 71 AU, but overpredicts
by 9\% the decrease in acceleration between intervals I and III 
of the Pioneer 10 observations.
(For comparison, the two different measurements of
the effect (SIGMA and CHASMP) themselves differ by 4\% in interval III.)
In any case, by accounting for the bulk of the acceleration, the
proposed mechanism makes it much more likely that the entire effect can be
explained without the need for new physics.

\end{abstract}

% PACS, the Physics and Astronomy Classification Scheme.
\maketitle

\section{INTRODUCTION}
\label{intro}

%************************2) PIONEER SPACECRAFT AND MISSION**********
In \cite{anderson}, Anderson {\it et al.}\ compare the measured trajectory
of several spacecraft against the theoretical trajectory computed from known
forces.  The find a small but significant
discrepancy, referred to as the unmodelled or anomalous acceleration.
It has an approximate magnitude of $\rm 8\times10^{-8}\;cm\;s^{-2}$ 
directed approximately towards the Sun.  Needless to say, {\it any} 
acceleration
of {\it any}  object that cannot be explained by conventional physics is
of considerable interest.
Explanations for this acceleration fall into two general categories - either
new physics is needed or some conventional force has been overlooked.

One of the most likely candidates for the anomalous acceleration is
non-isotropic radiation of spacecraft heat.  This is an appealing explanation
since the spacecraft dissipates about 2000 watts total; if only 58 watts
of this total power was directed away from the sun it could account for the
acceleration.  Several possible mechanisms have been debated in the
literature, but none are totally satisfactory.

In this paper we re-examine each proposed mechanism, explicitly 
including their time dependence.  We propose several additional mechanisms -
asymmetric RHU heat, misdirected feed radiation, and mis-modelled solar
reflectivity.
Finally, we compare the acceleration induced by the proposed mechanisms
with the measured data, and get reasonable agreement over the whole data span.

\section{The Anomalous Acceleration}

As the Pioneer spacecraft receded from the sun, solar forces decreased
and only gravitational forces, and an occasional maneuver, affected the
trajectory of the spacecraft.  Anderson, {\it et al.}\ noticed that a
small additional acceleration needed to be added to the known forces
to make the measured data and computations match.  This is the anomalous
acceleration, which started to become noticeable about 5 AU from the
sun, and was roughly the same for Pioneer 10 and 11. The onset is shown 
in Figure \ref{fig:EarlyAnom}.

Further constraints come from the ongoing study of Pioneer 10, 
where there are fewer confounding effects and
the data span is long enough to provide significant constraints due to
the radioactive decay of the heat sources.  
Figure \ref{fig:theory_vs_practice},
reproduced from \cite{Turyshev}, shows the measured acceleration 1987 to 1998.
(Although they have different horizontal axes, 
Figure \ref{fig:theory_vs_practice}
largely follows Figure \ref{fig:EarlyAnom} chronologically.  
Pioneer 10 was at 40 AU in 1987.)
The authors divide the 1987-1998 Pioneer 10 history into three intervals.
Interval I is January 1987 to July 1990, interval II from July 1990 
to July 1992, and interval III is
from July 1992 to June 1998.  The authors make this distinction
by looking at the spin rate of the craft - in intervals I and III 
it was decreasing smoothly, but in interval II it decreased quickly 
and irregularly.  They therefore consider the data from interval II 
to be less reliable than intervals I and III,
since whatever affected the spin (probably gas leaks) may also have affected
the acceleration.

More recent analyses have refined these results somewhat, though
the main conclusions remain unchanged.  
Table \ref{SummaryTable} shows the most recent results from \cite{anderson01},
which fits a constant, independent acceleration in each interval.
Accelerations are in units of $10^{-8}\;cm\;s^{-2}$.
SIGMA and CHASMP are two different and largely independent
trajectory modelling programs;
the difference between the programs is our best estimate of the real
uncertainties since it is far greater than the formal errors.
This data, taken at face value, shows that 57 directed watts 
can account for the acceleration in 1998, 
and that a 3\% decrease was observed between interval I and interval III.
\begin{table}[ht]
\begin{center}
\caption{Weighted Least Squares (WLS) results from 
Anderson, {\it et al.}\cite{anderson01}, and equivalent directed power for
a 241 kg spacecraft mass}
\label{SummaryTable}
\begin{tabular}{|c||c|c||c|c|} \hline
Interval& SIGMA & watts. & CHASMP & watts\\
	& accel.&        & accel. & \\
\hline \hline
Jan 87- Jul 90  & 8.00 $\pm$ 0.01 & 57.8 & 7.84 $\pm$ 0.01 & 56.7 \\ \hline
Jul 92 - Jul 98 & 8.25 $\pm$ 0.03 & 59.6 & 7.91 $\pm$ 0.01 & 57.2 \\ \hline
\end{tabular}
\end{center}
\end{table}
\section {PREVIOUS WORK}
\label{pioneer}

Many paper\cite{piodoc} and web\cite{pioweb} descriptions of the
Pioneer spacecraft are available.
In this section
we summarize the existing literature on the hypothesis that 
non-isotropic radiation is responsible for the unmodelled acceleration.

%************ Previous work *************************************

Murphy\cite{murphy} (and a related proposal by Scheffer\cite{Scheffer}) 
suggests that the  anomalous acceleration seen in the
Pioneer 10/11 spacecraft can be, ``explained, at least in part, by
non-isotropic  radiative cooling of the spacecraft.''
Katz\cite{katz} proposes that at least part of the acceleration is
generated by radiation from the RTGs reflecting off the back of the antenna.
Slusher (as credited by Anderson) proposed that the forward
and backward surfaces of the RTGs may emit non-equally.
Anderson, {\it et al.}\ argue in reply\cite{usmurphy,anderson01b,anderson01}
that none of these proposed sources adequately account for the acceleration.
\section{Discussion}

We consider asymmetrical radiation from 4 sources - the RTG heat (direct
radiation and reflection off the antenna),
the electrical power dissipated by the spacecraft,
the
radioisotope heater units (RHUs) on the spacecraft, and radiation from the
feed that misses the antenna.  We also consider one modelling error, a
mis-estimation of the reflectivity of the antenna to solar radiation.
The available power from all these sources changes in time.  In the following
discussion, let $d$ be the date, in years.
The sunward side of the spacecraft is the back, and the anti-sunward side,
in the direction of motion, is the front\cite{rearfront}.  We calculate
thrust in units of watts of directed (anti-sunward) radiation.
\subsection{Radiation of spacecraft power}
First, consider thermal
radiation from the body of the spacecraft.
A thought experiment shows that the electrical power
dissipated in the spacecraft must result in thrust.
The simplest model consists of
the main compartment as a 60 watt isotropic radiator, and the back of the 
antenna a mirror.  
The antenna subtends 120 degrees as seen from
the instrument compartment, so if the emitted radiation is isotropic,
the antenna intercepts 1/4 of the total radiation, and reflects it away
from the sun.  Since the main compartment is centered behind the antenna,
and since the sides, if anything, are worse radiators than the front, we
conclude that at least 25\% of spacecraft electrical power must be converted 
to thrust.

A more detailed analysis shows the radiation is even more anisotropic
than these arguments would suggest.
Assuming a uniform internal temperature and closed louvers, 
the power emitted from each 
surface is proportional to the area times the ``effective''
emissivity of the surface\cite{EffectiveEmissivity}. 
The sides and the rear of the compartment are covered with
multi-layer insulation(MLI)\cite{piodoc}, with an
effective emissivity of 0.007 to 0.01 \cite{Stimpson}.
The lowest emissivity material on the front
of the spacecraft is the surface of the louvers, with a emissivity of
0.04\cite{piodoc}.  Since the sides and the front have comparable
surface areas, then about 80\% of the total power will be
radiated though front (though hard to characterize heat leaks could reduce
this value).
Frontal radiation would be expected to be about 66\% efficient, 
assuming Lambertian emission.
Defining $\epsilon_{BUS}$ as the fraction of main compartment heat that 
is converted to thrust, we then expect $\epsilon_{BUS}$
to range between 0.25 (blockage arguments) to
0.52 (differential emissivity).

From \cite{anderson01}, the total electrical power is modelled
$$E(d) = (68+2.6~(1998.5-d))~\rm watts$$
and the thrust (assuming an 8 watt radio beam) is
$$ BUS(d) = \epsilon_{BUS} ~ (E(d)-8.0~{\rm watts}) $$
\subsection{Feed pattern of the radio beam}
An ideal radio feed antenna would illuminate its dish uniformly, 
with no wasted energy
missing the dish.  However, the feed is physically small and cannot create
such a sharp edged distribution, so some radiation always spills over the edge.
This radiation is converted 
to thrust with an efficiency of 1.7 since it directly subtracts from the
sun directed power and adds anti-sun power at a 
roughly 45 degree angle to the spin axis.  This produces thrust
 $$ RADIO(d) = \epsilon_{FEED}~(8~{\rm watts})~1.7$$ where
$\epsilon_{FEED}$ is the fraction of RF power that misses the antenna.
Since dish area is wasted if not fully illuminated, an optimum feed (for
transmission) will result in $\epsilon_{FEED} \approx 0.1$.

\subsection{Radiation from the RHUs}

From diagram 3.8-1 in \cite{piodoc}, 10 1-watt (in 1972) radioisotope heater units are
mounted to to external components (thrusters and the sun sensor) to keep them 
sufficiently warm.  The diagram
is not very specific, but the units to which they are mounted are primarily
behind the main dish.
Radiation from these components will contribute thrust, which we
model as 
$$ RHU(d) = \epsilon_{RHU} ~(10.0 ~{\rm watts}) 2^{-(d-1972)/88}$$  where
$\epsilon_{RHU}$ is the proportion of RHU heat converted to thrust.  Reasonable
values for $\epsilon_{RHU}$ might range from 0.0 to 0.5, with the latter
corresponding to components behind the dish radiating uniformly.

\subsection{Radiation from the RTGs}

The RTGs might contribute to the acceleration by radiating more to the front 
of the spacecraft than the rear, and/or by having their heat reflected
asymmetrically from the spacecraft.
The RTGs radiate all the thermal power that is not turned into
electricity, so $$RTG\_HEAT(d) = (2580~{\rm watts}) 2^{-(d-1972)/88}-E(
d)$$
In \cite{anderson01}, direct radiation asymmetry is estimated to contribute 
to thrust with an efficiency of at most $\pm 0.003$.
RTG reflection by the antenna was proposed by Katz, but argued against by
Anderson, primarily on the grounds that the RTGs are on-axis as seen
by the antenna.  We re-examine this argument here.  From figure 3.1-2 of
\cite{piodoc}, we see that
the centerline of the RTGs is behind the center of the antenna.  Measurements
from this diagram indicate this distance is about 23.8 cm.
Figure 3.1-3
of \cite{piodoc} shows the far end of the RTGs is 120.5 inches (or 3.06
meters) from the centerline.  From this geometrical data we can estimate
the area blocked by the antenna from each RTG\cite{scheffer01b}.
Numerical integration of these areas,
assuming Lambertian emission by the RTGs,
shows about 0.6\% of the near RTG radiation
and 0.4\% of the far RTG radiation fall upon the dish.
This energy is turned into thrust by two effects.  First, the antenna 
shadows radiation which
would otherwise go forward.  An angle in the middle of the antenna is about
17 degrees forward; this corresponds to an efficiency of 0.3 (the true
efficiency is probably higher since the edge is both at a greater angle
and more brightly illuminated.)
Next, the energy that hits the antenna must
go somewhere.  Some will be absorbed and re-radiated; some will bounce into 
space, and some
will bounce and hit the instrument compartment, and be reflected or re-radiated
from there.  A detailed accounting seems
difficult, but an overall efficiency of 0.7-0.9 seems reasonable (0.3 for shadowing
and 0.4-0.6 for reflection and re-emission).

We model the total thrust from RTG heat as
$$RTG(d) = \epsilon_{RTG}~RTG\_HEAT(d)$$
where $\epsilon_{RTG}$ is the proportion of RTG heat converted to thrust.
Combining the effects of this section, we expect $\epsilon_{RTG}$ to range
from 0.004 to 0.012.

\subsection{Antenna solar reflectivity}
\label{subsec:reflectivity}

The trajectory analysis programs fit the reflectivity of the spacecraft to
solar radiation, $\mathcal{K}$, as a force that falls off as $1/r^2$, where
$r$ is the heliocentric distance.  This fit can hide an otherwise
unmodelled acceleration.  Over a short time period,
during which $r$ varies little, any constant radial
acceleration can be absorbed into $\mathcal{K}$.  Over a longer period
of time the fitting procedure will mask any component of anomalous
acceleration that varies as $1/r^2$ and is less than the acceleration
corresponding to the allowed variation in $\mathcal{K}$.  In particular
the acceleration proposed in this paper will be partially masked since
it decreases with time and hence has a $1/r^2$ component.

The fitted solar reflectivity constant also provides a natural 
explanation for the onset of the anomalous acceleration.
Consider the case where the acceleration (from any cause) exists 
for all $r$.
When $r$ is small,
the fitting programs absorb the extra acceleration by adjusting the
value of $\mathcal{K}$.
As $r$ increases, the power available from this
source decreases, and eventually $\mathcal{K}$ runs into 
the limits allowed in the fit.
(Physically reasonable values perhaps range from 1.5 to about 1.9; they are
certainly greater than 1.0 and less than 2.0).  Once the limit of
adjustment for $\mathcal{K}$ is reached, it becomes constant and can
no longer mask the acceleration, which appears as shown in
figure \ref{fig:EarlyAnom}.  
It might be possible to see additional signs of this process in 
archival data - it would show up as a decrease in the fitted value of 
$\mathcal{K}$ as the spacecraft receded from the sun.

In this paper, we model the effect of any error in $\mathcal{K}$ 
by introducing a fictitious
force, whose value is simply the solar force on the
spacecraft times the error in $\mathcal{K}$.
We assume the distance from the sun, measured in AU, increases linearly from 20
AU in 1980 to 78.5 AU in 2001:
$$ r(d) = 20 + (d-1980)/21\cdot (78.5-20) $$
The thrust, in watts, is
$$ SOLAR(d) = K_{SOLAR}~\pi (1.37 ~{\rm m})^2 f_\odot/{r^2(d)} $$
where $f_\odot=1367 ~{\rm W/m}^{2}$(AU)$^2$
is the ``solar radiation constant'' at 1 AU, and   
$K_{SOLAR}$, the amount by which the solar reflection constant is underestimated.

\section{Comparison with experiment}

To compare the hypothesis with experiment, 
we sum the individual sources, then convert to acceleration by dividing by 
$c$, the speed of light, and $m$, the spacecraft mass (here 241 kg):
\begin{eqnarray*}
acc(d)&=&\frac{1}{c\cdot m} [RHU(d)+RTG(d)+\\& &RADIO(d)+BUS(d)-SOLAR(d)]
\end{eqnarray*}

We then compare with the plots from \cite{anderson01,Turyshev}.
The proposed explanation has 5 adjustable parameters.
In theory all are separable
since they decay at different rates; in practice the data are not good
enough to separate them and many fits are plausible.
One reasonable fit over the entire data span has the following coefficients:
$\epsilon_{RHU} = 0.5$,
$\epsilon_{RTG} = 0.0108$,
$\epsilon_{FEED} = 0.1$,
$\epsilon_{BUS} = 0.35$, and
$K_{SOLAR} = 0.3$.  

\begin{figure}[ht]
 \begin{center}
 \noindent
 \resizebox{8.65cm}{!}{
  \includegraphics[trim=1.2cm 0.4cm 0cm 0cm, clip=true]{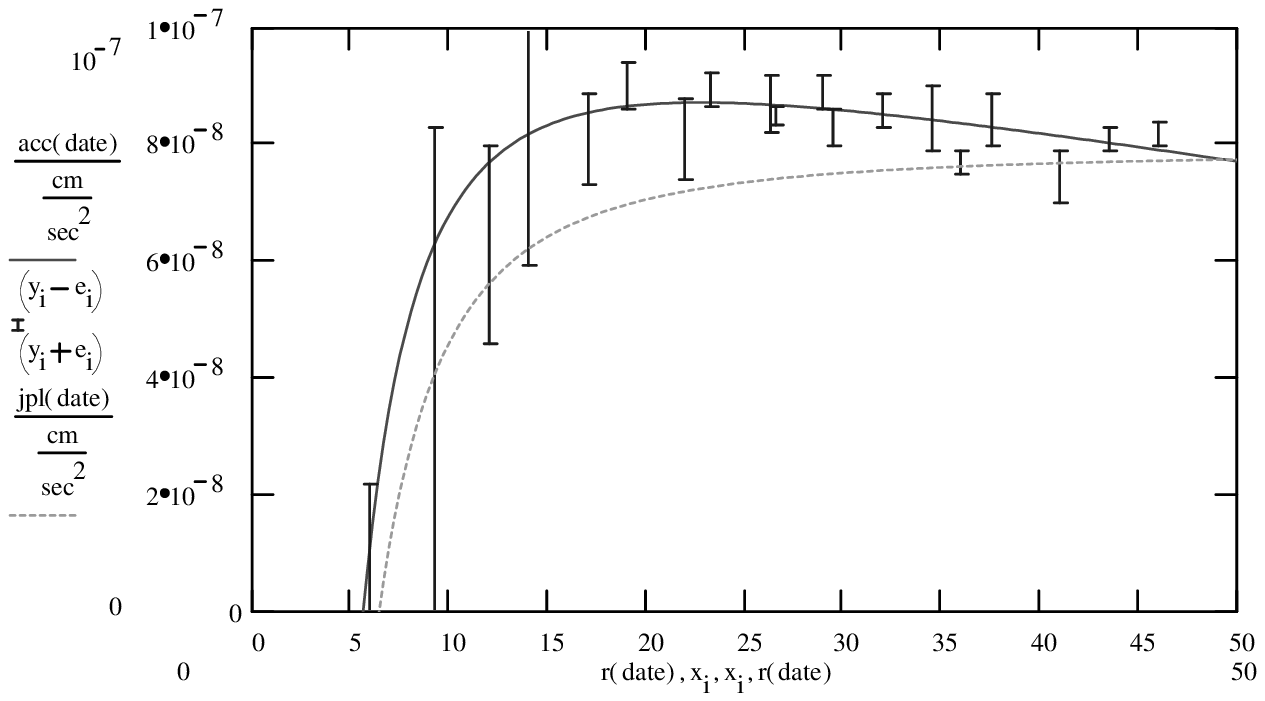}}
   \caption{Anomalous acceleration vs heliocentric distance in AU, from
   \cite{anderson01} (error bars),  model prediction
   from this paper (solid line), and constant acceleration plus
   reflectivity modelling error(dashed line)
    \label{fig:EarlyAnom}}
 \end{center}
\end{figure}

\begin{figure}[ht]
 \begin{center}
 \noindent
 \resizebox{8.65cm}{!}{\includegraphics{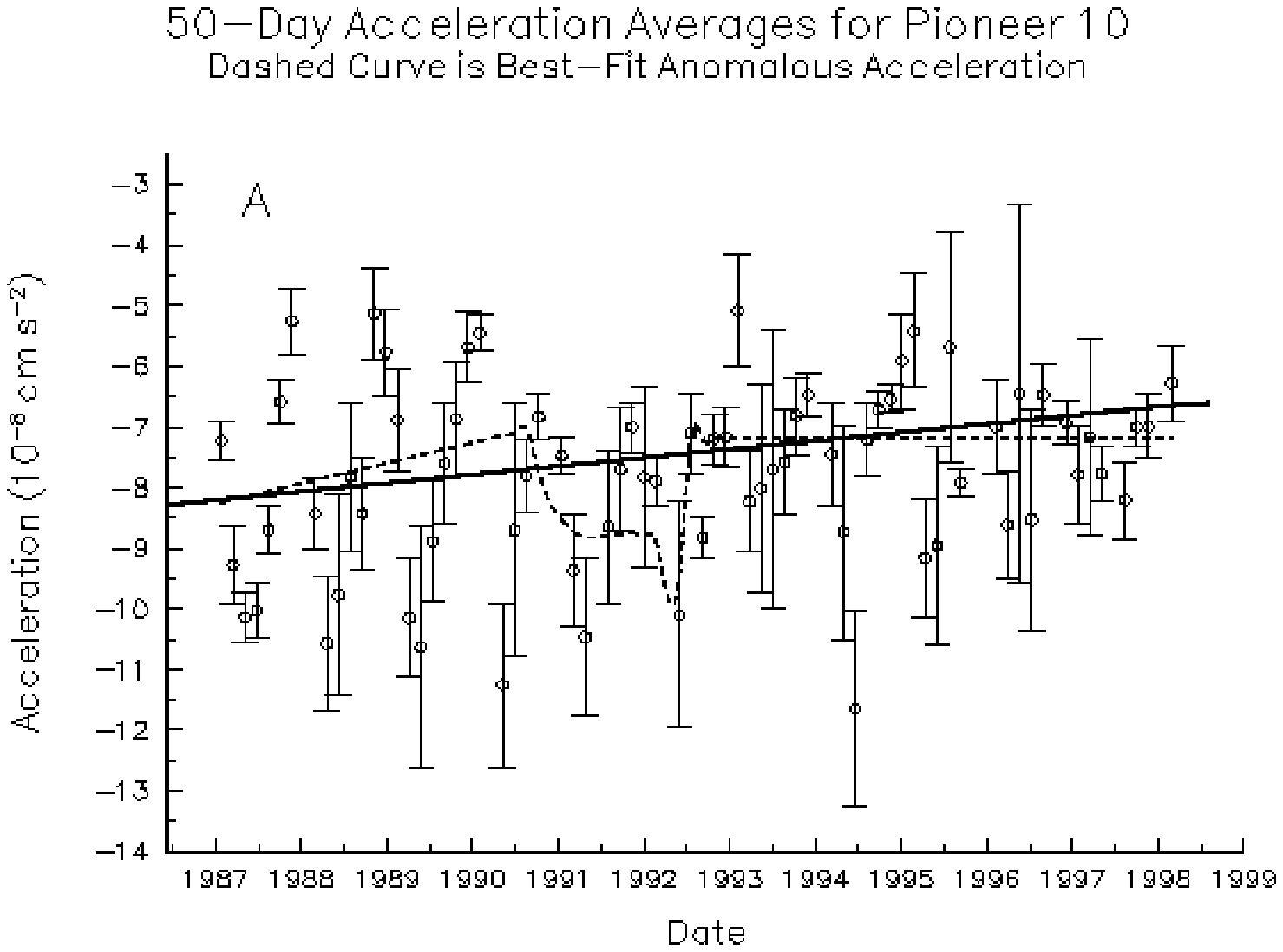}}
   \caption{Figure from \cite{Turyshev}, with fitted data added.
   The dotted line is Turyshev's empirical fit; the solid line is the
   model hypothesized in this paper.
    \label{fig:theory_vs_practice}}
 \end{center}
\end{figure}

This fit to the data is shown in Figures \ref{fig:EarlyAnom} and 
\ref{fig:theory_vs_practice}.  
The agreement seems reasonable in both regimes, and  
the proposed model provides a better fit to the early data
than the constant acceleration of \cite{anderson01}, even assuming
reflectivity mismodelling to account for the onset of the acceleration.
The fit from 1987 to 1998 also looks acceptable, as shown in Figure
\ref{fig:theory_vs_practice}.

Finally, we compare with the most recent results\cite{anderson01} that fit
a constant acceleration in each interval of the later Pioneer 10 data.  
The proposed model gives
an average thrust of 57.8 watts in interval I, and 51.0 watts in
interval III.  We can normalize the result to get the correct overall
average, or the right acceleration in interval I, but in either
case we would expect to see an 11.8\% decrease from
interval I to III, where only a 3\% decrease is observed.
The two different measurements of
the effect (SIGMA and CHASMP) themselves differ by 4\% in interval III.
If we treat this difference as a statistical result (a procedure of
dubious merit, but the best we can do) then the 9\% discrepancy is 2.25
standard deviations out.  This makes it unlikely at about the 2\% level
that this hypothesis alone accounts for all the measured result.

We can get a better fit (1.75 sigma) to the Pioneer 10 data by assigning 
different efficiencies to instrument heat and main compartment heat, 
at the cost
of an extra parameter and the need to consider instrument power dissipation
in detail\cite{scheffer01b}.

\section{Conclusions and future work}

There is surely an unmodelled effect on the Pioneer spacecraft, based upon
its thermal characteristics.  Rough estimates show
it can account for the magnitude of the unmodelled acceleration to within 
the errors, but overpredicts the rate of change.
In any case, the proposed explanation, by accounting for the bulk of the 
effect, makes it more likely that conventional physics can account for 
the entire unmodelled acceleration.
Conventional explanations for the remaining discrepancy
include other unmodelled effects such as
gas leaks, inaccuracies in the simple thermal model, or
the effects of
a complex fitting procedure applied to noisy data.
   
This explanation also explains some other puzzles: the values of acceleration
of Pioneer 10 and 11 would be expected to be similar, but not identical,
as observed. The acceleration would not have a strong effect on the spin;
most of the radiation will generate little torque.
Other spacecraft, built along the same general principles,
would be expected to show a similar effect, but planets and other large
bodies would not, as is observed.

More detailed modeling, using the Pioneer materials, construction details,
and history, might confirm or refute the proposed hypothesis, and
additional tracking could be useful as well.  However, such improvements
are limited since accurate thermal modelling is difficult\cite{anderson01}
and the spacecraft was not designed for this purpose.
Longer term,
other proposed experiments such as LISA\cite{LISA} are designed specifically
to reduce non-gravitational systematics (by a factor of about $10^5$) and
allow frequent and accurate tracking 
(a differential distance measurement, each second, accurate to $10^{-9}$ cm)
Assuming the anomalous acceleration exists at all heliocentric distances,
(as argued in section \ref{subsec:reflectivity}), then it should
be detectable in just a few seconds of LISA data.  On the other hand, 
if no unmodelled acceleration is detected in these more precise experiments, 
then almost surely the anomalous acceleration of Pioneer 10/11
is caused by overlooked prosaic sources such as those proposed here.

\section{Acknowledgements}

I'd like to thank Edward Murphy and 
Jonathan Katz for comments, suggestions, and helpful documents; 
Larry Lasher and Dave Lozier
answered questions about Pioneer.  John Anderson
suggested adding the statistical likelihood calculations.

%

%********************************************


\begin{thebibliography}{99}
\bibitem{EffectiveEmissivity}
When calculating radiation from multi-layer insulation, 
the correct value to use is the ``effective'' emissivity,
$ \epsilon_{eff}$,
which accounts for the lower temperature of the outer layer\cite{Stimpson}.
(Anderson\cite{anderson01b} 
points out that the
outer layer of the MLI has an emissivity of 0.70 according to \cite{piodoc}.
This is not a contradiction because of the temperature difference between the
interior of the spacecraft and the outer layer of the MLI.)
\end{thebibliography}
\end{document}